\documentclass[%
aip,
amsmath,amssymb,
tightenlines,
floatfix,
reprint
]{revtex4-1}

\usepackage{graphicx}
\usepackage{dcolumn}
\usepackage{bm}

\usepackage[utf8]{inputenc}
\usepackage[T1]{fontenc}

\usepackage{newtxtext,newtxmath}
\usepackage{microtype}

\usepackage{booktabs}
\usepackage{mathtools}
\usepackage{braket}

\usepackage[cal=dutchcal]{mathalpha}

\usepackage[svgnames]{xcolor}
\definecolor{cset-aps-blueberry}{RGB}{28,128,158}
\definecolor{cset-aps-blue}{RGB}{46,44,184}
\definecolor{cset-aps-turquoise}{RGB}{0,67,88}
\definecolor{cset-aps-limegreen}{RGB}{190,219,67}
\definecolor{cset-aps-green}{RGB}{31,138,112}
\definecolor{cset-aps-yellow}{RGB}{255,225,25}
\definecolor{cset-aps-orange}{RGB}{253,116,0}
\definecolor{cset-aps-red}{RGB}{219,0,43}

\definecolor{cset-aps-kobalt-medium}{RGB}{62,54,222}
\definecolor{cset-aps-kobalt-dark}{RGB}{28,24,150}

\definecolor{cset-aps-my-label-red}{RGB}{202,0,17}
\definecolor{cset-aps-my-label-blue}{RGB}{53,71,140}
\definecolor{cset-aps-my-label-gray}{RGB}{145,145,145}

\usepackage{hyperref}
\hypersetup{%
    colorlinks=true,
    linkcolor={cset-aps-red},
    linkbordercolor={cset-aps-red},
    filecolor={cset-aps-orange},
    filebordercolor={cset-aps-orange},
    citecolor={cset-aps-kobalt-medium},
    citebordercolor={cset-aps-kobalt-medium},
    urlcolor={cset-aps-kobalt-dark},
    urlbordercolor={cset-aps-kobalt-dark},
    menucolor={cset-aps-limegreen},
    menubordercolor={cset-aps-limegreen},
    breaklinks=true,
    pdfborderstyle={/S/U/W 2},
    pdfpagemode=UseOutlines,
    pdfstartpage={1},
}

\newcommand{\orcid}[1]{\href{https://orcid.org/#1}{\includegraphics[width=7pt,height=7pt]{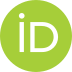}}}

\newcommand{\affULM}{Institut f{\"u}r Quantenphysik and Center for Integrated Quantum Science and Technology (IQST), Universit{\"a}t Ulm, Albert-Einstein-Allee 11, D-89081 Ulm, Germany}
\newcommand{\affHAN}{Institut f{\"u}r Quantenoptik, Leibniz Universit{\"a}t Hannover, Welfengarten 1, D-30167 Hannover, Germany}
\newcommand{\affTUDa}{Technische Universit{\"a}t Darmstadt, Fachbereich Physik, Institut f{\"u}r Angewandte Physik, Schlossgartenstr. 7, D-64289 Darmstadt, Germany}

\newcommand{\ie}{i.\,e.,}
\newcommand{\eg}{e.\,g.,}
\newcommand{\dd}{\text{d}}

\begin{document}

\preprint{AIP/123-QED}

\title{Optimal baseline exploitation in vertical dark-matter detectors \\ based on atom interferometry}
\author{Fabio Di Pumpo\,\orcid{0000-0002-6304-6183}}
\email{fabio.di-pumpo@uni-ulm.de, fabio.di-pumpo@gmx.de}
\author{Alexander Friedrich\,\orcid{0000-0003-0588-1989}}
\affiliation{\affULM}
\author{Enno Giese\,\orcid{0000-0002-1126-6352}\,}
\affiliation{\affTUDa}
\affiliation{\affHAN}

\collaboration{This article has been published as part of the \\\emph{{Large Scale Quantum Detectors Special Issue}} in \href{https://doi.org/10.1116/5.0175683}{AVS Quantum Science \textbf{6}, 014404 [2024]}}

\begin{abstract}
Several terrestrial detectors for gravitational waves and dark matter based on long-baseline atom interferometry are currently in the final planning stages or already under construction.
These upcoming vertical sensors are inherently subject to gravity and thus feature gradiometer or multi-gradiometer configurations using single-photon transitions for large momentum transfer.
While there has been significant progress on optimizing these experiments against detrimental noise sources and for deployment at their projected sites, finding optimal configurations that make the best use of the available resources is still an open issue.
Even more, the fundamental limit of the device's sensitivity is still missing.
Here we fill this gap and show that (a) resonant-mode detectors based on multi-diamond fountain gradiometers achieve the optimal, shot-noise limited, sensitivity if their height constitutes 20\,\% of the available baseline; (b) this limit is independent of the dark-matter oscillation frequency; and (c) doubling the baseline decreases the ultimate measurement uncertainty by approximately~65\,\%.
Moreover, we propose a multi-diamond scheme with less mirror pulses where the leading-order gravitational phase contribution is suppressed, compare it to established geometries, and demonstrate that both configurations saturate the same fundamental limit.
\end{abstract}

\maketitle

\section{\label{sec:Intro}Introduction}
Observations of departures from general relativity at the scale of galaxies, as evident in galaxy rotation curves, the dynamics of galactic clusters, or even on cosmic scales in the cosmic microwave background, strongly suggest~\cite{Bertone2018,Tulin2018} the presence of both dark matter (DM) and dark energy.
As of now, both have evaded all our direct detection efforts.
For DM, potential candidates can be described~\cite{Damour1994,Alves2000,Buckley2015,Jackson2023} as additional quantum fields that extend the Standard Model.
Any candidate might thus couple to (leptonic and baryonic) matter and hence effectively to atoms~\cite{Damour2010,Graham2013b} through their constituents.
As atoms are by definition quantum objects, they are a platform to implement high-precision quantum sensing protocols while at the same time being sensitive to DM~\cite{Safronova2018}.
Typically these technologies rely on superpositions of internal states or atomic trajectories, as illustrated by atomic clocks~\cite{Ludlow2015} and atom interferometers~\cite{Tino2021}.

Terrestrial atom-interferometric detectors with long baselines~\cite{Zhou2011,Schlippert2020,Asenbaum2020,Abend2023} are expected to complement existing approaches to DM searches\cite{Roszkowski2018} based on direct~\cite{Schumann2019} creation\cite{Buchmueller2017} or annihilation\cite{Arguelles2021} of dark matter which have proven unfruitful up until now.
These new instruments have been proposed in both~\cite{Zhan2020} horizontal~\cite{Canuel2018,Canuel2020} and vertical configurations~\cite{Badurina2020,El-Neaj2020,Abe2021} in synergy~\cite{Arvanitaki2018} with gravitational-wave detectors.
First demonstrator experiments which can already improve our constraints on dark matter~\cite{Abe2021,Badurina2020} are currently under construction~\cite{Mitchell2022}.
These large-scale quantum sensors become feasible through the suppression of common-mode noise between (at least) two atom interferometers, each probing DM at distinct spacetime locations.
Enhancing their sensitivity to both DM and gravitational waves can be achieved by expanding the number of atom-light interaction points~\cite{Graham2013}.
Consequently, large-momentum-transfer techniques are one strategy that can be used for an optimization of the signal.
The planned implementations frequently rely on (optical) single-photon transitions~\cite{Ludlow2015,Hu2017,Hu2020,Rudolph2020,Bott2023}, which offer the added advantage of suppressing laser phase noise~\cite{Yu2011,Graham2013}.

Possible detector sites are being currently evaluated~\cite{Junca2019,Canuel2018,Mitchell2022,Carlton2023,Arduini2023} with a focus on their noise characteristics.
In fact, gravity-gradient and other Newtonian noise has been identified as an important issue in classical terrestrial gravitational-wave detectors~\cite{Hughes1998,Beker2012,Harms2015Terrestrial}.
Similarly, these noise sources pose a major challenge for atom-interferometric DM and gravitational-wave detectors~\cite{Junca2019,Mitchell2022,Arduini2023}, especially for detecting DM in the sub-Hertz regime~\cite{Badurina2023}.
The dominant contribution of gravity gradients in a differential gradiometer setup can be suppressed by a so called two-diamond (or figure-of-eight or butterfly) geometry, whereas a single atom interferometer with such a geometry constitutes a gradiometer on the scale of the arm separation~\cite{Clauser1988,Marzlin1996,Kleinert2015,DiPumpo2023}.
In our article, we generalize this concept to a configuration encompassing multiple diamonds, where the role of both interferometer arms is interchanged between subsequent diamonds.
Whereas such three-diamond configurations have been demonstrated to suppress spurious Sagnac phases~\cite{Hogan2011atomic,Schubert2019}, an even number of diamonds has a similar effect on the dominant gravitational phase contributions.
A multi-diamond scheme in combination with periodically-shaped test masses is also key to proposals~\cite{Chiow2018} aiming at the detection of dark energy.
The proposed geometry also differs from configurations encompassing multiple diamonds where the role of both arms is not interchanged between subsequent diamonds.
Such schemes have been shown to enhance the signal of gravitational-wave detectors if the interrogation time of the interferometer matches the frequency of the wave~\cite{Graham2016resonant}.
This resonant-mode enhancement can also be observed in DM detectors~\cite{Arvanitaki2018,Badurina2022}, where sensitivity increases with the number of diamonds in the interferometer sequence.

An alternative strategy for noise suppression and, consequently, sensitivity enhancement involves incorporating multiple atom interferometers along the sensor's baseline~\cite{Chaibi2016}.
In this context, the quantity of interferometers and their distribution across the baseline can be fine-tuned and tailored to the specific environmental conditions, providing additional means to create versatile setups~\cite{Badurina2023}.

In contrast to these approaches, our article outlines how to exploit a vertical baseline in two different schemes of a multi-diamond gradiometer configuration, namely a geometry where the role of both arms remains the same between subsequent diamonds~\cite{Arvanitaki2018,Badurina2022} and a geometry where their role is interchanged.
While the optimal dimensions may vary depending on the specific characteristics of the prevailing noise, our considerations show that sensors for coherent DM waves, limited by shot noise, should allocate 20\,\% of the baseline to each atomic-fountain height to attain peak sensitivity. 
Given that shot noise represents the fundamental limit for such detectors, the demonstrated sensitivity saturates the ultimate limit of such an experiment.

Furthermore, this sensitivity bound is independent from the frequency of the DM oscillation, \ie{} its mass, and is dictated by the DM energy density confined within the volume of the detectors baseline.
Doubling the detector's baseline results in an approximate 65-\% reduction in the uncertainty associated with DM measurements.
Finally, we find that these results are identical for both schemes.

\section{Interferometer phase induced by dark matter}
We begin by deducing the principal DM contribution to the phase of a Mach-Zehnder atom interferometer whose atom-optical operations are performed via single-photon transitions~\cite{Ludlow2015,Hu2017,Hu2020,Rudolph2020,Bott2023}.
A generalization to large momentum transfer is discussed in Sec.~\ref{sec:MultiGrad}.
Our simple example serves to clarify and underscore the operational principles of this sensor type, without introducing excessive theoretical complexities.
In this spirit, we model ultralight scalar DM~\cite{Jackson2023} by a classical field.

Atoms are manipulated by atomic beam splitters or mirrors implemented via optical single-photon transitions between two internal atomic states.
The energy gap between both states corresponds to the atomic transition frequency $\Omega$ and is perturbed by the coupling of the atoms to the DM background.
Based on~\cite{Tulin2018} galactic observations and the assumed DM velocity distribution, we consider DM momentum as negligible and model the field as a position-independent plane wave with long coherence time.
Neglecting the spatial dependence of DM implies~\cite{Badurina2022} a mass range where the wavelength of the DM field is negligible on the length probed by the envisioned atom gradiometer.
While dropping this assumption is in principle possible, it leads to an additional potential probed by the sensor~\cite{DiPumpo2022}.
Even though such additional potentials may cause further phase contributions, similar to kinetic and gravitational ones~\cite{Derr2023}, the dominant phase originates in the rest mass-energy of the atom and is strictly associated with the internal transition frequency.
Similarly, a time-dependent modulation~\cite{Geraci2016,Derr2023} of the local gravitational acceleration due to a dressing of Earth's mass by the DM field may be included and used as an indicator for remnants of DM.
However, we assume in the following that this contribution is not the dominant one, which seems justified considering the respective energy scales~\cite{Derr2023}.

The interaction with the DM wave causes the atomic transition frequency to oscillate at frequency $\omega$, given by the mass of DM, which takes the form~\cite{Arvanitaki2018,Badurina2022}
\begin{equation}
    \Omega(t) = \Omega + \bar{\varepsilon} \delta \Omega  \cos{\left(\omega t+\phi\right)} .
\end{equation}
Here, $\bar{\varepsilon}\geq 0$ is the dimensionless coupling of both internal states to DM~\cite{Derr2023}.
This quantity is the parameter that is measured or bounded by the DM detector.
Beyond pure phenomenology, it may be linked to coupling parameters of fundamental particles and atomic constituents that do not depend on the specific atomic species involved~\cite{Safronova2018}.

The phase $\phi$ of the DM wave is in general a free parameter and can vary between different shots.
The oscillation amplitude of the atomic transition frequency~\cite{Hees2016,Filzinger2023}
\begin{equation}
\label{eq:deltaOmega}
    \delta \Omega = \Omega \frac{m_\text{P}c^2}{\hbar \omega}\sqrt{8 \pi\frac{ \varrho_\text{DM} L_\text{P}^3 }{m_\text{P} c^2}} 
\end{equation}
can be connected to the local energy density $\varrho_\text{DM}\cong 0.4$\,GeV/cm${}^3$ of DM and its frequency $\omega$.
The energy scale is given the Planck mass $m_\text{P}$ and the characteristic volume is given by the cube of the Planck length $L_\text{P}= \hbar/ (m_\text{P}c)$.
Here, $c$ denotes the speed of light and $\hbar$ Planck's constant.

In addition to a modification of the atomic transition frequency, a coupling to DM affects the center-of-mass motion of the atom.
In principle, this interaction introduces additional phases~\cite{Geraci2016} in atom interferometers, since they represent inertial sensors and are by that routinely employed as accelerometers~\cite{Graham2016}.
However, the dominant contribution arises from the clock phase originating from the time intervals during which the atom is in different internal states~\cite{Arvanitaki2018} along each interferometer arm.
Hence, we neglect effects from these sub-dominant and higher-order couplings to DM in the following.

\begin{figure}[h]
    \centering
	\includegraphics{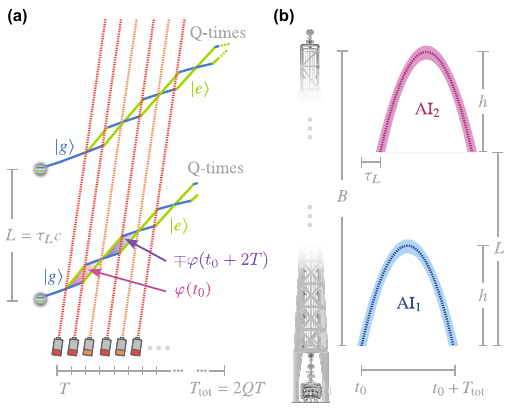}
    \caption{{\sffamily\bfseries (a)} 
    Spacetime diagram of two multi-diamond gradiometer configurations consisting of two interferometers separated by a distance $L=c\tau_L$ and generated by single-photon pulses (dotted, red and orange) sent at appropriate multiples of the interrogation time $T$. 
    The first scheme ($-$) is generated by the pulses indicted in red, while the additional  $\pi$ pulses shown in orange are only present in the second ($+$) scheme.
    The overall duration of the interferometer $T_\text{tot}=2QT$ scales with the number of diamonds $Q$.
    We indicate the alternating sequence of ground state $\ket{g}$ (blue) and excited state $\ket{e}$ (green) of the atomic clouds during the interferometer along their respective trajectories.
    The phase difference acquired during one diamond $\varphi(t_0)$ depends on the initial time $t_0$ and is identical but shifted in time in the second ($+$) scheme.
    In the first ($-$) scheme it alternatingly flips its sign in subsequent diamonds, since the role of both arms is interchanged.
    {\sffamily\bfseries (b)}
    Spatial extension $h$ and midpoint trajectory (dashed) of the two atomic fountains that are used as multi-diamond interferometers $\text{AI}_1$ (blue) and $\text{AI}_2$ (red).
    They are separated by a distance $L$ distributed on the baseline $B$ of the detector and their start (and end) is delayed by a time $\tau_L=L/c$ originating in the propagation of the light between the two atomic ensembles.
    The additional extension originating from the atomic recoil and subsequent wave-packet propagation is illustrated by the shaded area surrounding the respective midpoint trajectories.%
    \footnote{Adapted VLBAI model in {\scriptsize\sffamily\bfseries FIG.~1.~(b)} courtesy of the VLBAI team of the Institute of Quantum Optics, Leibniz University Hannover.}
    }
    \label{fig:setup}
\end{figure}

Due to the inherent time symmetry between the branches of the Mach-Zehnder sequence making up the first diamond of the atom interferometer, shown in Fig.~\ref{fig:setup}\,(a), the unperturbed clock phase cancels.
Such a sequence consists of a $\pi/2$ pulse that coherently splits the atomic beam, a $\pi$ pulse that reflects both branches, and a second $\pi/2$ pulse that interferes them.
The cancelation only arises when all three pulses are separated by equal interrogation times $T$.

The coupling to DM breaks this inherent symmetry by inducing a time-dependent atomic transition frequency.
Hence, any remaining clock phase can serve as a direct probe for the coupling of DM to the atom provided all additional phases resulting \eg{} from motional effects or the finite speed of light on the scale of the interferometers branch separation, are either of higher order or can be sufficiently well characterized.
Under these assumptions, the induced phase difference
\begin{align}
\begin{split}
\label{eq:phase_MZI}
    \varphi(t_0)=- \bar{\varepsilon}\delta \Omega\left[\int\limits_{\mathrlap{t_0}}^{\mathrlap{t_0+T}}{\dd t\cos \left(\omega t+\phi\right)}-\int\limits_{\mathrlap{t_0+T}}^{\mathrlap{t_0+2T}}{\dd t\cos{\left(\omega t+\phi\right)}}\right]
\end{split}
\end{align}
acquires a dependence on the initial time $t_0$ of the Mach-Zehnder sequence.
The sign flip in the second half of the interferometer stems from the interchanged role of both internal states after the action of the $\pi$ pulse at time $t_0+T$.

While this phase contains a signature of DM, other, more dominant contributions can arise, \eg{} from the motion of the atom in external potentials like gravity present in terrestrial setups.
However, a differential setup as sketched in Fig.~\ref{fig:setup}\,(a) serves to isolate the phase from Eq.~\eqref{eq:phase_MZI}.
This procedure is discussed in Sec.~\ref{sec:MultiGrad} after generalizing the scheme to multiple diamonds instead of a plain Mach-Zehnder interferometer.

\section{Multi-diamond Gradiometer Signal}\label{sec:MultiGrad}
So far we have demonstrated that the phase of a Mach-Zehnder interferometer is susceptible to DM oscillations.
We now discuss two possible generalizations to multi-diamond schemes that enhance the sensitivity:
(i) In a first scheme ($-$), the second $\pi/2$ pulse that interferes both branches is omitted.
To keep the distance between them sufficiently small and subsequently overlap them at the end of the light-pulse sequence, the atom is periodically redirected by subsequent $\pi$ pulses separated by a period $2T$, as shown in  Fig.~\ref{fig:setup}\,(a) by the dotted red lines.
This procedure creates a geometry with additional Mach-Zehnder diamonds after the first Mach-Zehnder interferometer.
After a total duration of $T_\text{tot}= 2QT$, where $Q$ describes the number of diamonds, a second $\pi/2$ pulse finally interferes both branches.
(ii) The second scheme ($+$) redirects both arms whenever they begin to cross each other by introducing additional $\pi$ pulses shown in  Fig.~\ref{fig:setup}\,(a) by orange dotted lines.
As a consequence, all pulses are separated by the interrogation time $T$. 
Such a configuration has been demonstrated to resonantly enhance the signal of gravitational-wave detectors~\cite{Graham2016resonant} by a factor of $Q$ and has been adapted for the detection of DM~\cite{Arvanitaki2018,Badurina2022}.

Similarly to the symmetry in a single Mach-Zehnder interferometer, the role of the arms is interchanged between two subsequent diamonds in the first scheme.
In this case, the sign of the acquired phase also flips~\cite{Chiow2018}, while in the second scheme the role of both arms is not reversed, so that there is no change of sign.
As a consequence, we find for $Q$ diamonds a phase contribution for the first ($-$) and second ($+$) scheme
\begin{align}
\begin{split}
    \label{eq:multi-diamond}
    \Phi_\mp(t_0)=\sum\limits_{q=1}^{Q}\left(\mp 1\right)^{q-1}\varphi\left(t_0+2(q-1) T\right),
\end{split}
\end{align}
which depends on the initial time $t_0$ of the multi-diamond sequence.
While Eq.~\eqref{eq:multi-diamond} describes the dominant phase induced by DM, the overall phase is still sensitive to other and possibly larger phase contributions as well as their associated noise.

For this reason, one usually resorts to common-mode operation in a gradiometer-type configuration~\cite{Hu2020}.
In these configurations common light pulses drive the transitions of two atom interferometers separated by a distance $L$, distributed along a baseline $B$ of the detector.
The situation is shown in Fig.~\ref{fig:setup} for a vertical configuration with atom-fountain interferometers.
In this case, the differential phase $\delta \Phi_{\mp} = \Phi_{\mp}(t_0+\tau_L)-\Phi_{\mp}(t_0)$ removes the dominant inertial phases as well as most of the noise, even though gravity-gradient noise is expected~\cite{Junca2019,Mitchell2022,Arduini2023} to be a severe but ultimately solvable~\cite{Chaibi2016} challenge.
Here, the first ($-$) scheme suppresses leading-order gravitational phases and the associated noise, while the second ($+$) scheme does not feature such an intrinsic symmetry.
The delay time $\tau_L=L/c$ that arises from the propagation of the light between the atom interferometers is crucial for the scheme:
only due to the finite propagation speed of light the DM field is probed at two different instances in time.

\begin{figure}
    \centering
	\includegraphics{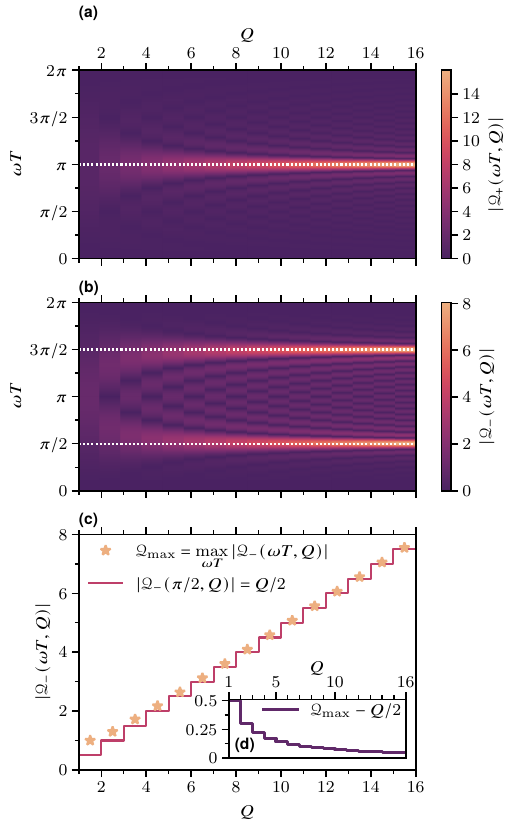}
    \caption{%
   Panels {\sffamily\bfseries (a)} and {\sffamily\bfseries (b):} Visualization of the interrogation-mode functions $\mathcal{Q}_{\mp}(\omega T,Q)$ of a DM sensor with $Q$ diamonds based on the second scheme ($+$) or the first scheme ($-$) and its dependence on the product $\omega T$ of DM frequency $\omega$ and half-diamond duration $T$, respectively.
   For the second scheme ($+$) the interrogation mode function $\mathcal{Q}_{+}$ always attains its maximum $Q$ at $\omega T=\pi$.
   In case of the first scheme ($-$) the interrogation-mode function $\mathcal{Q}_{-}$ attains its peak value close to the horizontal dashed lines determined by $\omega T = \pi/2$ or odd integer multiples thereof with increasing $Q$.~Panel {\sffamily\bfseries (c)}: Optimization of the signal amplitude of a $Q$-loop multi-diamond DM sensor by maximizing the interrogation-mode function $\mathcal{Q}_{-}(\omega T,Q)$ with respect to $\omega T$ at fixed $Q$.
   In comparison to the simplified case of choosing $\omega T = \pi/2$ with $\left|\mathcal{Q}_{-}(\pi/2,Q)\right|=Q/2$, we obtain a deviation of the maximal value $\mathcal{Q}_\text{max}=\max_{\omega T} |\mathcal{Q}_{-}(\omega T,Q)|$ from $Q/2$ in the limit of small $Q$.
   The exact deviation $\mathcal{Q}_\text{max}-Q/2$, obtained by numerical optimization, is plotted in the inset {\sffamily\bfseries (d)} of panel {\sffamily\bfseries (c)} as a function of $Q$.
   It attains the peak deviation $0.5$ for the Mach-Zehnder interferometer with $Q=1$.
   Consequently, in this case one can double the interrogation-mode function and hence the signal amplitude by performing such an optimization.
   However, starting at $Q\gtrsim 5$ this additional benefit rapidly vanishes and the interrogation-mode function becomes indistinguishable from the now dominating $Q/2$ scaling originating from the resonant-mode enhancement for larger $Q$.}
    \label{fig:interrogation-mode_fktn}
\end{figure}

However, the phase $\phi$ of the DM field will change from shot to shot.
Consequently, it is only possible to measure the amplitude of the stochastic background induced by DM on the differential phase, \ie{} the sensor detects $\Phi_\text{S}=\big[2\int_0^{2\pi}{\dd\phi \,\delta \Phi^2_{\mp}/(2\pi)}\big]^{1/2}$.
This signal amplitude takes the form
\begin{equation}
\label{eq:SignalAmpl}
    \Phi_\text{S}=\bar{\varepsilon} \frac{8\delta \Omega}{\omega} \left|\sin \frac{\omega\tau_L}{2}  \mathcal{Q}_{\mp}(\omega T, Q)\right|
\end{equation}
and is determined by the interrogation-mode functions
\begin{subequations}
\label{eq:interrogation-mode}
    \begin{equation}
\label{eq:interrogation-mode+}
    \mathcal{Q}_+(\omega T, Q) =  \frac{1}{2} \sin (Q \omega T) \tan \frac{\omega T}{2},
\end{equation}
which is in agreement with previous treatments~\cite{Badurina2022}, and
\begin{equation}
\label{eq:interrogation-mode-}
    \mathcal{Q}_-(\omega T, Q) =   
    \begin{cases}
     \sin^2 \frac{\omega T}{2} \cos \left(Q\omega T\right) / \cos \omega T & \text{for } Q\text{ odd} \\
     \sin^2 \frac{\omega T}{2} \sin \left(Q\omega T\right) / \cos \omega T & \text{for } Q\text{ even}
    \end{cases}
\end{equation}
\end{subequations}
that includes all dependence on $T$ and $Q$.
In particular, the interrogation-mode function $\mathcal{Q}_{-}$ differs for odd and even numbers of diamonds, as visualized in Fig.~\ref{fig:interrogation-mode_fktn}\,(b), while  $\mathcal{Q}_{+}$ shows no such behavior as shown in Fig.~\ref{fig:interrogation-mode_fktn}\,(a).
This distinction is a direct consequence of the factor $\left(-1\right)^{q-1}$ in the sum of Eq.~\eqref{eq:multi-diamond} that arises because the role of both interferometer arms is interchanged between two subsequent diamonds.

For a Mach-Zehnder interferometer with $Q=1$, we find $\mathcal{Q}_{\mp} = \sin^2( \omega T/2)$.
In this case, the signal amplitude reduces to the expression also found for large-momentum-transfer single-diamond setups~\cite{Arvanitaki2018,Carlton2023}.
The respective multi-diamond signal of the second ($+$) scheme in gravitational-wave detectors~\cite{Graham2016resonant} also leads to the interrogation-mode function given by Eq.~\eqref{eq:interrogation-mode+}. 
Moreover, the DM signal for this configuration has been derived previously~\cite{Badurina2022}.

To study resonant-mode enhancement~\cite{Graham2016resonant}, we observe the relations
\begin{equation}
    | \mathcal{Q}_-(\pi/2, Q) | = Q/2  \text{ and } | \mathcal{Q}_+(\pi, Q) | = Q
\end{equation} 
and hence we refer in the following to $\omega T = \pi /2$ and $\omega T = \pi$, respectively, as the resonant mode, where the signal amplitude is amplified by the number of diamonds.
In this case, the interrogation time $T$ is adapted to the frequency $\omega$ of the DM field. 
For ultralight DM of any given mass, one can find such a time, only being limited by experimental constraints.
While the maxima differ by a factor of two, we show here that the optimal sensitivity is the same for both schemes.

However, the choice of $\omega T = \pi /2$ does not necessarily maximize the signal for small $Q$, as shown in Fig.~\ref{fig:interrogation-mode_fktn}\,(c).
In fact, the maximum of $| \mathcal{Q}_- |$ approaches $Q/2$ from above, even though already at $Q> 5$ the difference drops below $5~\%$ and soon becomes negligible.
So instead of discussing the optimal choice of $\omega T$ for each configuration, we simply resort to the resonant mode as defined above.
Note that in the other scheme the choice $\omega T = \pi $ is always optimal.

While noise might enter the signal differently for odd or even $Q$, as apparent from the two-diamond setup, also the DM signal changes due to the form of $\mathcal{Q}_-$.
This difference cannot be observed for resonant-mode detection.
However, for a sensor operated in far-off resonant detection mode with a total duration $T_\text{tot} = 2 Q T \ll 1/\omega$, the signal amplitude scales as $1/Q^2$.
As a consequence, a multi-diamond configuration operated in this mode has no benefit over the Mach-Zehnder setup.
This limit arises naturally if the peak sensitivity is not reachable within the experimentally available durations for the interferometer.
In this case, even numbers of diamonds are more strongly suppressed by increasing $Q$ than odd numbers, which can be directly seen from an expansion of Eq.~\eqref{eq:interrogation-mode-}.
Nevertheless, this regime is anyways not favorable for operation and we focus on resonant-mode detection in the following.

As a further simplification, we can assume that the phase of DM varies slowly on the time scale of the propagation delay, that is  $\omega \tau_L \ll1$, which is well justified for ultralight DM.
Expanding Eq.~\eqref{eq:SignalAmpl} to first order in $\omega \tau_L$, the signal amplitude reduces to the compact expression $ \Phi_\text{S}= \bar{\varepsilon} 4 \delta \Omega \tau_L  \left| \mathcal{Q}_\mp\right| $.

One of the techniques routinely applied to enhance the sensitivity of atom interferometers is the transfer of large momenta~\cite{Rudolph2020}.
In fact, most gravitational-wave and DM detectors are planned~\cite{Canuel2018,Schubert2019,Badurina2020,Mitchell2022} with a design that includes such large-momentum-transfer technologies as one key component.
An additional benefit of an even number of transferred momenta is that the atom travels in the same internal state for most of the interrogation time $T$, which lifts some of the requirements on the lifetime of the excited state.
Moreover, it suppresses differential phases that may arise for an atom being in different internal states on both branches, \eg{} induced by spatially varying Zeeman shifts.

Our treatment can be generalized to large-momentum-transfer schemes in analogy to other approaches~\cite{Schubert2019,Badurina2022,Carlton2023}.
In such a sequence the (even) number $N$ of transferred momenta further enhances the signal amplitude.
For $\omega \tau_L \ll1$ we find
\begin{equation}
\label{eq:LMTSignalAmpl}
    \Phi_\text{S}=  \bar{\varepsilon} 4 \delta \Omega  \tau_L N \left| \mathcal{Q}_{\mp}\right|
\end{equation}
as a generalization of Eq.~\eqref{eq:SignalAmpl}, so that also the case $N=1$ is correct.
Note that the second scheme ($+$) corresponds to the same pulse sequence used in previous treatments~\cite{Arvanitaki2018,Badurina2022}.
However, to our knowledge the first scheme ($-$) has not been discussed in this context.

There is in general a difference between the propagation delay $\tau_L=L/c$ that is given by the initial distance $L$ between both interferometers and the length $B$ of the baseline.
In addition one can consider the propagation delay between the position of the lasers and the lower interferometer as well as the propagation delay between the reflectors at the other end of the baseline and the atom interferometer.
However, even in this case, only $\tau_L$ enters the signal amplitude to the lowest order~\cite{Graham2016resonant,Badurina2022}.
Similarly, the light propagation delay is neglected on the scale of the distance between both branches of a single interferometer, also assumed for the case $N=1$ studied so far.

\section{Sensitivity to Dark Matter}
A measurement of the amplitude of the stochastic background induced by DM on the differential phase yields data to estimate the dimensionless DM-coupling parameter $\bar{\varepsilon}$.
However, all experiments are prone to noise, be it a fundamental one (like shot noise), introduced by experimental imperfections, or inevitable fluctuations like gravity-gradient noise.  
From the signal amplitude given in Eq.~\eqref{eq:LMTSignalAmpl} we find through Gaussian error propagation that the uncertainty of $\bar{\varepsilon}$ has the form
\begin{equation}
\label{eq:uncertainty}
    \Delta \bar \varepsilon = \frac{\Delta \Phi_S}{ 4 \delta \Omega  \tau_L N \left| \mathcal{Q}_\mp\right|}
\end{equation}
and depends on the fluctuations $\Delta \Phi_S$ of the signal amplitude.
By introducing the signal-to-noise ratio $\operatorname{SNR}= (\bar \varepsilon / \Delta \bar \varepsilon)^2$ as an estimator for the signal strength, we directly observe
\begin{equation}
\label{eq:SNR}
    \bar \varepsilon =\Delta \bar \varepsilon\sqrt{\operatorname{SNR}}  = \frac{\sqrt{\operatorname{SNR}} }{ 4 \delta \Omega  \tau_L N \left| \mathcal{Q}_\mp\right|}\Delta \Phi_S,
\end{equation}
which can also be derived via arguments about the power spectral density~\cite{Badurina2022}.
Equation~\eqref{eq:SNR} sets an upper limit on the coupling $\bar \varepsilon$ for a given $\operatorname{SNR}$ and experiment, and is usually estimated in theoretical studies~\cite{Badurina2023b}. 
Since it is directly proportional to $\Delta \bar \varepsilon$, we focus in the following discussion on its optimization. 

The explicit form of the fluctuations $\Delta \Phi_S$ will depend on the local environment, the specifics of the experimental realization and setup, as well as intrinsic constraints.
While it is possible to maximize the signal amplitude by operating in resonant mode, this choice might not be optimal to minimize $\Delta \bar \varepsilon$.
In fact, $\Delta \Phi_S$ will depend in many cases on the interrogation time $T$.
Thus, it is insufficient to maximize the interrogation-mode function $\left| \mathcal{Q}_{\mp}\right|$ alone.
Instead, the best strategy is to find an interrogation time where the fraction $\Delta \Phi_S/ \left| \mathcal{Q}_{\mp}\right|$ is minimized.
Since such a procedure strongly depends on the experimental site and a detailed noise analysis, such a full discussion is beyond the scope of this article.
However, we will derive a limit for the sensitivity to $\bar{\varepsilon}$ in the following that is imposed by the fundamental detection shot noise and optimize the exploitation of the baseline of the detector.

In a first step, we assume that $\Delta \Phi_S$ is independent of $T$.
In this case, the minimal uncertainty $\Delta \bar \varepsilon$ indeed arises for resonant-mode detection, where the signal amplitude is maximized by $| \mathcal{Q}_-|  = Q/2$ with $\omega T = \pi/2$ or $| \mathcal{Q}_+|  = Q$ with $\omega T = \pi$, respectively.
Even if there is a small dependence of  $\Delta \Phi_S$ on $T$, the following treatment may hold if the necessary conditions $| Q \dd \Delta \Phi_S /\dd T|_{\omega T=\pi/2}\ll 1$ or $| Q \dd \Delta \Phi_S /\dd T|_{\omega T=\pi}\ll 1$ are fulfilled.
Combined with $T_\text{tot}= 2 Q T$ we arrive at the uncertainty
\begin{equation}
    \Delta \bar \varepsilon = \frac{\pi}{2}\frac{\Delta \Phi_S  }{ N \delta \Omega \omega  \tau_L T_\text{tot}} 
\end{equation}
for both types of pulse sequences.
Even though the maxima for both schemes differ by a factor of one half, the duration of resonant-mode detection also differs by a factor of two, so that the value for $\Delta \bar \varepsilon$ is independent of the specific implementation.

Referring to the setup shown in Fig.~\ref{fig:setup}\,(b), this expression can be connected to the spatial dimensions of a vertical detector.
For optimal usage of the available resources and space, we assume that the vertical baseline $B= L + h$ is given by the initial separation $L$ of the atoms and the height $h$ of an individual atomic fountain, where the atoms are launched and imaged at the same location. 
Hence, this height can be roughly estimated by the midpoint trajectory following a parabola.
Instead of an atomic fountain with an initial launch, the treatment can also be adopted to describe an operation in drop mode, that could in principle allow for interleaved interferometers~\cite{Savoie2018}.
However, for a fountain setup the duration of the atom-interferometer sequence $T_\text{tot}\cong \sqrt{ 8 h /g} - 2 v_\text{r}/(g Q)$ can be connected to the fountain height and the velocity $v_\text{r}$ transferred by atomic recoil to the atoms during the light-pulses.
As a consequence, the gravitational acceleration $g$ enters for vertical, terrestrial detectors.
For simplicity, we assume that the deviation from the midpoint trajectory can be neglected for sufficiently large $Q$.
Moreover, the propagation delay $\tau_L = (B-h)/c$ depends on the dimensions of the baseline as well.

To optimize the exploitation of the baseline, we start by the restrictive assumption that $\Delta \Phi_S$ is not only independent of $T$, but in addition also independent of $T_\text{tot}$.
In this case, the uncertainty $\Delta \bar{\varepsilon}$ from Eq.~\eqref{eq:uncertainty} is minimized for a choice of $h=B/3$ and takes the optimal value
\begin{equation}
\label{eq:sesitivity}
    \Delta \bar{\varepsilon} =  \frac{3\sqrt{3\pi}\Delta \Phi_S}{32 N} \frac{m_\text{P} c^2}{\hbar \Omega} \frac{L_\text{P}}{R_\text{E}}\sqrt{ \frac{m_\text{E} c^2}{ \varrho_\text{DM}B^3 }} .
\end{equation}
Here, the radius $R_\text{E}$ of Earth and its mass $m_\text{E}$ enter through the gravitational acceleration $g$ at its surface.
Moreover, we have used the explicit form of $\delta \Omega$ given in Eq.~\eqref{eq:deltaOmega}.

The uncertainty still depends on the fluctuations of the signal amplitude $\Delta \Phi_S$, so that low noise is necessary for the determination of $\bar{\varepsilon}$ with high precision.
One can increase the sensitivity by choosing a larger atomic transition frequency $\Omega$, even though this direct proportionality is an artifact of our assumption that both internal states couple equally to DM.
The precision is enhanced by increasing the baseline of the interferometer, with a scaling behavior of $B^{-3/2}$, which is better than inversely proportional.
In particular, Eq.~\eqref{eq:sesitivity} highlights that the energy density of DM is effectively probed by a volume determined by the length of the baseline.
Furthermore, this limit of resonant-mode detection is independent of the frequency $\omega$ of the DM wave and by that independent of its mass.

In a next step, we make less restrictive assumptions on the noise and derive the fundamental limit on the sensitivity.
It is saturated if both atom interferometers are shot-noise limited and the integration time $T_\text{int}$ is smaller than the coherence time of the DM wave~\cite{Centers2021stochastic,Badurina2022}.
For a differential measurement, shot noise adds in and we find $\Delta \Phi_S = \sqrt{2/(\nu n_\text{at}) }$.
Here, $n_\text{at}$ is the number of atoms and the number of repetitions $\nu$ is connected to the integration time through $T_\text{int}= \nu T_\text{tot}$, assuming ideally a vanishing dead time between subsequent runs.
We therefore implicitly assume that the next run of the experiment can be prepared while the previous atom-interferometer sequence is ongoing.
In principle, one could also increase this factor by resorting to interleaved schemes.
In any case, if $\Delta \Phi_S $ is independent of $T$ but only depends on $T_\text{tot}$ we arrive at
\begin{equation}
    \Delta \bar \varepsilon = \frac{\pi  }{ \sqrt{2n_\text{at}} N \delta \Omega \omega  \tau_L \sqrt{T_\text{tot} T_\text{int}}} 
\end{equation}
for resonant-mode detection.
Assuming that the overall integration time is fixed and limited by long-term drifts of the detector, we find in analogy to the discussion above an optimal height $h=B/5$, \ie{} 20\,\% of the baseline.
This choice leads to the uncertainty
\begin{equation}
    \Delta \bar{\varepsilon} =  \frac{5}{64 N} \sqrt{\frac{10\pi}{n_\text{at} \nu } }\frac{m_\text{P} c^2}{\hbar \Omega} \frac{L_\text{P}}{R_\text{E}}\sqrt{ \frac{m_\text{E} c^2}{ \varrho_\text{DM}B^3 }},
\end{equation}
which constitutes the fundamental limitation for the sensitivity.
The height of the two atom fountains is adjusted in such a way that the baseline of the resonant-mode detector is exploited optimally.
This fundamental limit is independent of the mass of DM, \ie{} its frequency, and is suppressed by the number of repetitions and atoms used per run.
Again, we observe a scaling behavior of $B^{-3/2}$.

\section{Conclusions}
In this article, we derived a general expression for the uncertainty of the dimensionless constant $\bar{\varepsilon}$ that describes the coupling of both internal states of an atom to a (coherent) DM field~\cite{Derr2023}, as measured by an atomic multi-diamond gravimeter operated with single-photon transitions.
The key ingredient was the identification of an interrogation-mode function $\mathcal{Q}_{\mp}$ that highlights the benefits of operation in resonant mode. 
This result can be used to optimize the detector for specific noise characteristics, which in turn depend on the site of the detector.
In principle, the fraction $\Delta \Phi_S/ \left|\mathcal{Q}_{\mp}\right|$ of the fluctuations of the signal amplitude and the interrogation-mode function has to be minimized to find the optimal interrogation time $T$.
Such an optimum will generally depend on the frequency of the DM field and typical frequencies of the noise characteristics.

However, for shot-noise limited atom-fountain interfero\-meters, where the integration time is smaller than the coherence time of the DM wave, we derived an optimal fountain height of 20\,\% of the baseline.
The observed sensitivity for resonant-mode operation is independent of the mass of DM, \ie{} its frequency, and constitutes the ultimate limit of such a configuration.
In the future it could be further enhanced by relying on quantum-metrological techniques like squeezing or entangling the atomic input states~\cite{Anders2021,Greve2022} once these techniques have sufficiently matured for application in large-momentum-transfer interferometers. 
In this case, the scaling with respect to the atom number changes to $n_\text{at}^{-1}$ instead of $n_\text{at}^{-1/2}$.
However, the remainder of the expression is not affected and the optimal choice of the fountain height does not change.
Overall, we observe a scaling of the uncertainty with $B^{-3/2}$ so that doubling the baseline leads to a decrease in the uncertainty by roughly 65\,\%.
In fact, the energy density of DM is effectively probed by a volume determined by the length of the baseline.
The uncertainty $\Delta \bar \varepsilon$ of the coupling parameter has been obtained from Gaussian error propagation, so that $5\Delta \bar \varepsilon$ corresponds to a five-sigma discovery. 
With the help of Eq.~\eqref{eq:SNR} using five times the uncertainty, the value
\begin{equation}
    \bar \varepsilon_{5\sigma} = \frac{25 }{64 N} \sqrt{\frac{10\pi \operatorname{SNR}}{n_\text{at} \nu } }\frac{m_\text{P} c^2}{\hbar \Omega} \frac{L_\text{P}}{R_\text{E}}\sqrt{ \frac{m_\text{E} c^2}{ \varrho_\text{DM}B^3 }}
\end{equation}
has to be observed in a measurement.

Our results can be applied to different situations such as drop-mode operation instead of using a symmetric fountain setup where the atoms are launched and imaged at the same location so that the height is given by the apex of the trajectory.
While the free-fall time $T_\text{tot}$ and number of possible diamonds decreases in drop mode, an interleaved operation~\cite{Savoie2018} might be possible.
It would introduce a more favorable scaling behavior with the integration time, so that the optimization of the drop height will lead to a different result.
Moreover, one can generalize our study to multi-gradiometry~\cite{Badurina2023} with more than two atom interferometers placed in the baseline of the detector. 
In this case, not only the drop height, but also the spacing and the number of interferometers can be optimized.

Moreover, we have discussed two schemes where (i) the role of both arms changes between subsequent diamonds and (ii) the arms are redirected at the end of each diamond.
Both schemes have different interrogation times in resonant-mode detection, but lead to the same optimal sensitivity for the coupling parameter.
While the first scheme is intrinsically insensitive to the dominant gravitational phase contribution~\cite{Clauser1988,Marzlin1996,Kleinert2015,DiPumpo2023} for even $Q$ and therefore leads to a suppression of gravity-gradient noise by relying at the same time on less pulses, the second scheme has been already discussed in the context of DM or gravitational-wave detection~\cite{Graham2016resonant, Arvanitaki2018,Badurina2022}.

So far, we have neglected further corrections due to a nonnegligible recoil, which becomes more important with increasing numbers of large-momentum-transfer pulses.
In such situations, finite-speed-of-light effects on the scale of the distance between the interferometer branches might become relevant as well.
However, this separation decreases with an increasing number of diamonds, suppressing such effects.
This suppression therefore depends on the frequency of the DM field for resonant detection and for fixed $T_\text{tot}$.
Moreover, terrestrial, vertical long baselines require a significant amount of chirping in the laser frequency~\cite{Tan2017_Time} to compensate the Doppler detuning, which should be accounted for in a more detailed treatment.

For a comprehensive analysis, one can also include DM couplings that lead to phase contributions beyond the leading-order effects discussed in this article.
In particular, one can include the motion of the atoms~\cite{Geraci2016} as common for conventional gradiometers or gravimeters, where a perturbative operator approach~\cite{Ufrecht2020perturbative} seems to be well-suited to incorporate effects of DM on the atoms' motion.

In this article, we have used a scalar field described by plane waves to model DM.
This model can be extended to include effects like the presumed velocity distribution of DM, \ie{} a spatial dependence of the wave~\cite{DiPumpo2022}, as well as a finite coherence time of the DM wave~\cite{Centers2021stochastic,Badurina2022}.
Moreover, one can go beyond the limiting case of ultralight, scalar, and classical DM fields to observe differences in the signal amplitude depending on the particular candidate of DM~\cite{Bertone2018,Roszkowski2018}.

\begin{acknowledgments}
We are grateful to W. P. Schleich for his stimulating input and continuing support and to the VLBAI team at the Institute of Quantum Optics, Leibniz University Hannover, for providing us with a schematic of their VLBAI tower for use and adaptation in Fig.~\ref{fig:setup}\,(b).
We also thank D. Derr as well as the QUANTUS and INTENTAS teams for fruitful and interesting discussions.
Moreover, we thank L. Badurina and Ch. McCabe for their helpful insights on the second interferometer sequence~\cite{Graham2016resonant,Badurina2022}.
F.D.P. is grateful to the financial support program for early career researchers of the Graduate \& Professional Training Center at Ulm University and for its funding of the project ``Long-Baseline-Atominterferometer Gravity and Standard-Model Extensions tests'' (LArGE).
The Qu-Gov project in cooperation with ``Bundesdruckerei GmbH'' is supported by the Federal Ministry of Finance (Bundesministerium der Finanzen, BMF).
A.F. is grateful to the Carl Zeiss Foundation (Carl-Zeiss-Stiftung) and IQST for funding in terms of the project MuMo-RmQM.
The QUANTUS and INTENTAS projects are supported by the German Space Agency at the German Aerospace Center (Deutsche Raumfahrtagentur im Deutschen Zentrum f\"ur Luft- und Raumfahrt, DLR) with funds provided by the Federal Ministry for Economic Affairs and Climate Action (Bundesministerium f\"ur Wirtschaft und Klimaschutz, BMWK) due to an enactment of the German Bundestag under Grant Nos. 50WM2250D-2250E (QUANTUS+) and 50WM2177-2178 (INTENTAS).
E.G. thanks the German Research Foundation (Deutsche Forschungsgemeinschaft, DFG) for a Mercator Fellowship within CRC 1227 (DQ-mat).
\end{acknowledgments}

\section*{Author declarations}
\subsection*{Conflict of interest}

\noindent The authors have no conflicts to disclose.

\subsection*{Author contributions}
\noindent{\sffamily\small\textbf{Fabio Di Pumpo}} Conceptualization (supporting); Formal analysis (supporting); Methodology (supporting); Validation (equal); Visualization (supporting); Writing - original draft (supporting); Writing – review and editing (equal).
{\sffamily\small\textbf{Alexander Friedrich}} Conceptualization (supporting); Formal analysis (supporting); Methodology (supporting); Validation (equal); Visualization (lead); Writing - original draft (supporting); Writing – review and editing (equal).
{\sffamily\small\textbf{Enno Giese}} Conceptualization (lead); Formal analysis (lead); Methodology (lead); Validation (equal); Visualization (supporting); Writing - original draft (lead); Writing – review and editing (equal).

\section*{Data availability}
The data that support the findings of this study are available within the article.

\section*{References}
\bibliography{Literatur}

\end{document}